\documentclass[12pt]{article} % default is 10pt
\usepackage[utf8x]{inputenc} % to set input encoding (not needed with XeLaTeX)
\usepackage[T1]{fontenc} % to set font encoding in latex; difference noticed for <>
\usepackage{amsmath,amssymb,bbm,simplewick,xfrac,wasysym} % for mathematical symbols
\usepackage{sectsty} % to change title styles
\usepackage{color,xparse} % to use color and to link arXiv properly
\usepackage{cite} % for compact citations

\usepackage{geometry} % to change the page dimensions
\usepackage{fancyhdr} % This should be set AFTER setting up the page geometry
\usepackage{setspace} % to set spacing in the document
\usepackage[titles]{tocloft} % to alter the style of the ToC

\usepackage[unicode]{hyperref} % for bookmarks in Acrobat Reader and using hyperlinks
\hypersetup{bookmarksnumbered=true, bookmarksopen=true, bookmarksopenlevel=2, breaklinks=true, citecolor=blue, colorlinks=true, linkcolor=red, linktoc=page, pdfborder={0 0 0}, pdfkeywords={N=3 CSM, }{Harmonic Superspace, }{Wilson Loops}, pdfstartview=FitH, pdfauthor={YtH,DJain}, pdfsubject={Wilson Loops and Harmonic Superspace}, pdftitle={N=3 Harmonic Super-Wilson Loops}, plainpages=false, unicode=true, urlcolor=cyan}

\DeclareUnicodeCharacter{"0393}{\Gamma}
\DeclareUnicodeCharacter{"0394}{\Delta}
\DeclareUnicodeCharacter{"0398}{\Theta}
\DeclareUnicodeCharacter{"039B}{\Lambda}
\DeclareUnicodeCharacter{"039E}{\Xi}
\DeclareUnicodeCharacter{"03A0}{\Pi}
\DeclareUnicodeCharacter{"03A3}{\Sigma}
\DeclareUnicodeCharacter{"03A5}{\Upsilon}
\DeclareUnicodeCharacter{"03A6}{\Phi}
\DeclareUnicodeCharacter{"03A8}{\Psi}
\DeclareUnicodeCharacter{"03A9}{\Omega}
\DeclareUnicodeCharacter{"03B1}{\alpha}
\DeclareUnicodeCharacter{"03B2}{\beta}
\DeclareUnicodeCharacter{"03B3}{\gamma}
\DeclareUnicodeCharacter{"03B4}{\delta}
\DeclareUnicodeCharacter{"03B5}{\epsilon}
\DeclareUnicodeCharacter{"03B6}{\zeta}
\DeclareUnicodeCharacter{"03B7}{\eta}
\DeclareUnicodeCharacter{"03B8}{\theta}
\DeclareUnicodeCharacter{"03D1}{\vartheta}
\DeclareUnicodeCharacter{"03B9}{\iota}
\DeclareUnicodeCharacter{"03BA}{\kappa}
\DeclareUnicodeCharacter{"03BB}{\lambda}
\DeclareUnicodeCharacter{"03BC}{\mu}
\DeclareUnicodeCharacter{"03BD}{\nu}
\DeclareUnicodeCharacter{"03BE}{\xi}
\DeclareUnicodeCharacter{"03C0}{\pi}
\DeclareUnicodeCharacter{"03C1}{\rho}
\DeclareUnicodeCharacter{"03C3}{\sigma} %σ
\DeclareUnicodeCharacter{"03C4}{\tau}
\DeclareUnicodeCharacter{"03C5}{\upsilon}
\DeclareUnicodeCharacter{"03C6}{\phi}
\DeclareUnicodeCharacter{"03D5}{\varphi}
\DeclareUnicodeCharacter{"03C7}{\chi}
\DeclareUnicodeCharacter{"03C8}{\psi}
\DeclareUnicodeCharacter{"03C9}{\omega}
\DeclareUnicodeCharacter{"21D0}{\Leftarrow}
\DeclareUnicodeCharacter{"0212B}{\AA}
\DeclareUnicodeCharacter{"00B7}{\cdot}
\DeclareUnicodeCharacter{"266A}{\eighthnote}
\DeclareUnicodeCharacter{"266B}{\twonotes}
\PrerenderUnicode{ä}

\definecolor{title}{rgb}{0.1,0.5,0.9}
\definecolor{abst}{rgb}{0.366,0.366,0.266}
\definecolor{sect}{rgb}{0,0,0}
\definecolor{ssect}{rgb}{0,0,0.0}
\definecolor{sssect}{rgb}{0.3,0.3,0.3}
\definecolor{appsect}{rgb}{0.5,0.8,0.1}
\definecolor{ref}{rgb}{0.0,0.0,0}
\newcommand{\Title}[1] {\title{\color{title}\Huge #1}}
\newcommand{\TPheader}[3]
{\date{}\maketitle\thispagestyle{fancy}\pagenumbering{alph}\lhead{#1}\chead{#2}\rhead{#3}\cfoot{}}
\newcommand{\makepage}[1] {\newpage\pagenumbering{#1}}

\newcommand{\Abstract}[1] {\begin{abstract}\normalsize #1 \end{abstract}}

\sectionfont{\color{sect}\Large\bf}
\subsectionfont{\color{ssect}\large\bf}
\subsubsectionfont{\color{sssect}\normalsize\bf}
\renewcommand{\appendix}{\setcounter{section}{0}\sectionfont{\color{appsect}\Large\bf}\renewcommand{\thesection}{\Alph{section}}\renewcommand*{\theHsection}{app.\the\value{section}}} % Hsection command for correct hyperlinking in .pdf!

\newcommand\references[1] {\begin{thebibliography}{[99]}\addcontentsline{toc}{section}{References} #1
\end{thebibliography}}
\NewDocumentCommand\arXivid{m+g}{\IfNoValueTF{#2}{\href{http://arxiv.org/abs/#1}{\tt arXiv:#1}}{\href{http://arxiv.org/abs/#1}{\tt arXiv:#1{\small [#2]}}}}%\arXivid{hep-th/xxxx.xxxx}{hep-th}
\newcommand\cqg[4] {\href{https://doi.org/#4}{{\it Class.\ Quant.\ Grav.\ }{\bf #1} (#2) #3}}
\newcommand\jhep[4] {\href{https://doi.org/#4}{{\it JHEP\ }{\bf #1} (#2) #3}}
\newcommand\npb[4] {\href{https://doi.org/#4}{{\it Nucl.\ Phys.\ B\ }{\bf #1} (#2) #3}}
\newcommand\plb[4] {\href{https://doi.org/#4}{{\it Phys.\ Lett.\ B\ }{\bf #1} (#2) #3}}
\newcommand\prd[4] {\href{https://doi.org/#4}{{\it Phys.\ Rev.\ D\ }{\bf #1} (#2) #3}}

\newcommand\eqs[1] {\begin{align}#1\end{align}}
\newcommand\eqsn[1] {\begin{align*}#1\end{align*}}

\newcommand\eqst[1] {\begin{multline}#1\end{multline}}

\newcommand\eqsc[1] {\begin{gather}#1\end{gather}}

\newcommand\eqsg[1] {\eqs{\begin{aligned}#1\end{aligned}}}
\newcommand\equ[1] {\begin{equation}#1\end{equation}}

\renewcommand\i {\dot{\iota}}
\newcommand\half {\tfrac{1}{2}}
\newcommand\s {\sigma}
\renewcommand\( {\left(}
\renewcommand\) {\right)}
\DeclareMathOperator{\Tr}{tr}

\newcommand\D {{\cal D}}
\newcommand\F {{\cal F}}

\newcommand\N {{\cal N}} 
\renewcommand\O {{\cal O}}
\renewcommand\P {{\cal P}}
\renewcommand\S {{\cal S}}
\newcommand\W {{\cal W}}
\newcommand\bR {{\mathbb R}}

\newcommand\braket[1] {\left\langle#1\right\rangle}
\newcommand\tint {{\textstyle ∫}}
\newcommand\bs[1] {\boldsymbol{#1}}

\newcommand\nn {\nonumber\\}

\numberwithin{equation}{section} % Equations numbered as <section>.#
\begin{document}
\Title{$\N=3$ Harmonic Super-Wilson Loop}

\author{Yu-tin Huang$^{η_1,η_2 }$\footnote{\href{mailto:ythuang@phys.ntu.edu.tw}{ythuang@phys.ntu.edu.tw}}$~$ and Dharmesh Jain$^{ψ}$\footnote{\href{mailto:d.jain@saha.ac.in}{d.jain@saha.ac.in}}\bigskip\\
\emph{${}^{η_{1}}$Department of Physics, National Taiwan University,}\\ \emph{No.1, Sec.4, Roosevelt Road, Taipei 10617, Taiwan} \medskip\\ 
\emph{${}^{η_{2}}$Physics Division, National Center for Theoretical Sciences,}\\
\emph{National Tsing-Hua University,}\\
 \emph{No.101, Section 2, Kuang-Fu Road, Hsinchu 30013, Taiwan} \medskip\\ 
\emph{${}^{ψ}$Theory Division, Saha Institute of Nuclear Physics,}\\ \emph{1/AF Bidhan Nagar, Kolkata 700064, India}
}

\TPheader{\today}{}{NCTS-TH/1805} %\TPheader{Date}{...}{PREPRINT}

\Abstract{We study supersymmetric Wilson loops in $d=3$, $\N=3$ harmonic superspace, leading to a construction of a supersymmetrized generalization of the $\frac{1}{3}$-BPS Wilson loop for $\N=3$ gauge theories. This also includes a generalization of the $\frac{1}{6}$-BPS loop for ABJM theory. We perform a `one-loop' computation of the vacuum expectation value of this operator directly in superspace and compare with the known $\N=2$ localization results at large $N$. This comparison also lets us identify certain fermionic contributions that do not receive any subleading corrections.
}

\tableofcontents
\makepage{arabic}
%%%%%%%%%%%%%%%%%%%%%%%%%%%%%%%%%%%%%%%%%%%%%%%%

\section{Introduction}
The power of supersymmetry to simplify computations and gain insights cannot be overstated. It sheds light on hidden structures and illuminates relationships among seemingly different objects. A perfect example of this power is given by the Wilson loops/Scattering amplitudes duality in $d=4$, $\N=4$ super-Yang Mills (SYM) theory. Even though evidence for such a duality existed\cite{Drummond:2007cf,Drummond:2007bm,Bern:2008ap,Drummond:2008aq,Drummond:2008vq,Berkovits:2008ic}, only after the construction of a supersymmetrized Wilson loop (WL) in superspace \cite{Mason:2010yk,CaronHuot:2010ek} has the duality been confirmed for all helicity sectors of the amplitude. In three-dimensions, while similar evidence in the case of four-point amplitude/four-gon Wilson loop for $\N=6$ ABJM theory\cite{Aharony:2008ug} exists\cite{Chen:2011vv,Bianchi:2011dg}, extending beyond four-points immediately forces us into the remaining sectors (in terms of R-symmetry instead of helicity) of the theory. This motivates us to construct supersymmetric Wilson loops in superspace.  

After the introduction of ABJM theory, various Wilson loop operators with different amounts of preserved supersymmetry were studied extensively. Earlier efforts dealt with construction and perturbative computations of $\frac{1}{6}$-BPS WL \cite{Gaiotto:2007qi,Drukker:2008zx,Chen:2008bp,Rey:2008bh}. Localization was applied to evaluate the vacuum expectation value (vev) of this WL in \cite{Kapustin:2009kz} and the results were found to match the perturbative calculations at large $N$ limit. $\half$-BPS operators were constructed later in \cite{Drukker:2009hy} and more calculations followed in \cite{Bianchi:2013zda,Bianchi:2013iha} where even finite $N$ contributions were computed. Being `cohomologically equivalent' to the $\frac{1}{6}$-BPS operator, the localization results do not differ for these two operators. In \cite{Ouyang:2015bmy}, a classification was given for Wilson loops preserving various amounts of supersymmetry in $\N=2,⋯,6$ Chern-Simons (CS) matter theories. New Wilson loops in $\N=4$ theories have been constructed recently in \cite{Mauri:2017whf}.

In this ever-expanding literature of construction, classification and computation involving Wilson loops, we present here a supersymmetrization of the simplest WL operator in three-dimensional CS matter theories including ABJ(M) theories. Such an attempt has been made in \cite{Rosso:2014oha} for ABJM theory in the framework of `ordinary' $\N=6$ superspace. It was also pointed out that there are at least three reasons why such a WL cannot be dual to the scattering amplitudes of ABJM theory. The main issue is the non-chiral nature of the superspace that leads to torsion, which does not allow a straightforward identification of the kinematics on the two sides of the duality\cite{Beisert:2012gb}. So we content ourselves with the `well-studied' framework of $\N=3$ harmonic superspace \cite{Buchbinder:2009dc, Buchbinder:2008vi} to construct the supersymmetrized Wilson loop\footnote{Harmonic Superspace was originally constructed for $d=4$, $\N=2$ supersymmetric theories in \cite{Galperin:1984av}.}. This is to have as much manifest (off-shell) supersymmetry as possible along with a notion of chirality (or `harmonic analyticity') built-in. 

In the next section we consider a warm-up exercise of constructing a supersymmetrized WL in $\N=2$ superspace and a sample localization computation. Then we review the $d=3$, $\N=3$ harmonic superspace in Section \ref{Sc:Rev} before constructing the supersymmetrized $\frac{1}{3}$-BPS WL in Section \ref{Sc:SWL}. This leads to a generalization for $\frac{1}{6}$-BPS WL in ABJ(M) theories. In Section \ref{Sc:Com}, we compute perturbatively the `one-loop' vev of this new WL operator directly in harmonic superspace. Finally, we compare the perturbative result with localization computation and comment on future outlook in Section \ref{Sc:com}.

\section{Warm-up}\label{Sc:Wu}
We construct here a supersymmetrized Wilson loop operator in $d=3$, $\N=2$ superspace with coordinates $\{x^μ(x^{(αβ)}), θ^α,\bar{θ}^α\}$, where the vector index $μ=0,1,2$ and spinorial index $α=1,2$ correspond to the $SO(2,1)≃SL(2,\bR)$ group\footnote{The vector $x^μ$ can be traded for a real second-rank symmetric tensor $x^{αβ}≡x^μ(γ_μ)^{αβ}$ with the help of $d=3$ ``gamma''-matrices. We do not need the explicit basis but the relation $x^{αβ}x_{αβ}=-2x^μx_μ≡-2|x|^2$ will be quite useful to know.}. Though it is rather straightforward, we think this analysis has not appeared in the literature in this form so we discuss it as a warm-up exercise leading to the less trivial $\N=3$ superspace in the next section.

The $\N=2$ supersymmetry algebra has the following set of gauge-covariant superspace derivatives: $\{\D_μ(\D_{(αβ)}),$ $\D_α,\bar{\D}_α\}$. These satisfy the following algebra:
\eqsg{\{\D_α,\bar{\D}_β\} &=\i \D_{αβ} +ε_{αβ}W\,;\qquad \{\D_α,\D_β\}=0 \\
[\D_α,\D^{βγ}] &=δ_α^{(β}\bar{W}^{γ)} \\
[\D_μ,\D_ν] &=\i\F_{μν}\,.
\label{N2alg}}
The Jacobi identities give further relations among the field strengths $W, W^α$ and $\F_{μν}$. One such relation is $\D_αW=-\i \bar{W}_α$ along with the chirality constraint $\D_α\bar{W}_β=0$ \cite{Nishino:1991sr,Gates:1991qn}.

The supersymmetrization of the familiar $\half$-BPS Wilson loop in chiral superspace then looks like\equ{\W(x,θ,\bar{θ})=\frac{1}{\dim R}\Tr_R\P\exp∫dτ\left[\tfrac{-\i}{2}\dot{x}_A^{αβ}A_{αβ} +\dot{θ}^αA_α +|\dot{x}_A|W \right]\equiv\frac{1}{\dim R}\Tr_R\P e^{w},
\label{N2WL}}
where $x_A^{αβ}=x^{αβ}+\i θ^{(α}\bar{θ}^{β)}$. We can do the component analysis of the connections and field strengths, leading to the fields of $\N=2$ vector multiplet $\{a_{αβ},\s,λ_α,\bar{λ}_α,D\}$ along with the field strength $f_{αβ}$:
\eqsc{W_|=\s\,;\quad \D_αW_|=\bar{λ}_α\,;\quad \bar{\D}_αW_|=λ_α\,;\quad \D_α\bar{\D}_βW_|=f_{αβ} +ε_{αβ}D \nn
{\bar{\D}_{(α}A_{β)}}_|=a_{αβ}\,;\quad {\bar{\D}·A}_|=\s\,;\quad {\bar{\D}^2A_α}_|=λ_α\,;\quad {\D_α\bar{\D}·A}_|=\bar{λ}_α\,;\quad {\D^α\bar{\D}^2A_α}_|=D \nn
{A_{αβ}}_|=a_{αβ}\,;\quad {\D^αA_{αβ}}_|=\bar{λ}_β\,;\quad {\bar{\D}^αA_{αβ}}_|=λ_β\,;\quad {\D^α\bar{\D}^βA_{αβ}}_|=D\,.
}
Here ${}_|$ denotes that all $θ$'s are set to vanish. Also relevant is ${\D_{(α}\bar{\D}^2A_{β)}}_|=f_{αβ}$. It is now trivial to verify that the $θ$-independent piece of the exponent in \eqref{N2WL} reduces to the well-known bosonic expression: $∫dτ(\i\dot{x}^μA_μ+|\dot{x}|\s)$.

It can be easily checked that the $\W(x,θ,\bar{θ})$ preserves `some' supersymmetry:
\eqs{δ\W≡ε^γ\D_γ\W(x,θ,\bar{θ}) &\sim \Tr_R\P\left\{e^{w}∫dτ\big(ε^γ\D_γ\big[\tfrac{-\i}{2}\dot{x}_A^{αβ}A_{αβ} +\dot{θ}^αA_α +|\dot{x}_A|W \big]\big)\right\} \nn
%&\propto ∫dτ\big(-\i ε^γ\dot{x}_A^{αβ}\F_{γ,αβ} +\cancelto{\scriptstyle 0}{ε^γ\dot{θ}^α\F_{γ,α}} \;\;+|\dot{x}_A|ε^γ\D_γW\big) \nn
&\sim\Tr_R\P\left\{e^{w} ∫dτ\(\i ε^α\(\dot{x}^A_{αβ} +|\dot{x}^A|ε_{αβ}\)\bar{W}^β\)\right\}  \,.
\label{N2Wvar}}
In arriving at the last step, we have used the algebra \eqref{N2alg} to convert covariant derivatives acting on connections into the corresponding field strengths, and terms that look like field-dependent gauge transformations of the connections, i.e. $\dot{x}^{A,αβ}\D_{αβ}(ε^γA_γ)$, are dropped as $\W(x,θ,\bar{θ})$ is gauge invariant. The BPS condition for the purely bosonic WL requires $x^μ(τ)$ to be an infinite line in Minkowski space or a great circle on $S^3$ and one can choose it to satisfy $|\dot{x}|=1$\cite{Kapustin:2009kz,Ouyang:2015bmy}. Since \eqref{N2Wvar} for the supersymmetrized case results in a similar equation, we will also consider $|\dot{x}^A|=1$. This does not determine $θ(τ)$ completely but only up to a function of $τ$: $θ(τ)=f(τ)θ_0,\,\bar{θ}(τ)=f^{-1}(τ)\bar{θ}_0$.\footnote{It is most likely that one needs to consider superconformal transformations of the WL operator to fully determine the $θ(τ)$ profile consistent with the circular bosonic WL. We do not pursue this exercise here. Thus, we will not evaluate the $τ$-integrals explicitly and leave all the $τ$-dependence of the coordinates intact.} Hence, constant solutions for $ε$ can still be found for these configurations, where the condition $ε^α\(\dot{x}^A_{αβ} +|\dot{x}^A|ε_{αβ}\)=0$ projects half of the degrees of freedom, thus preserving two real degrees of freedom, i.e. $\frac{1}{2}$-BPS. 

Given the Lagrangian and propagators of \cite{Nishino:1991sr,Gates:1991qn}, one should be able to compute the vev of the WL \eqref{N2WL} perturbatively in superspace as well as in components at different $θ$-orders. However, we will skip this analysis here and comment on the non-perturbative analysis instead. Using the localization results of \cite{Kapustin:2009kz} where a $\N=2$ theory on $S³$ (of radius $r$) is considered, we can obtain an `exact' result for the vev of the supersymmetrized Wilson loop. Since the path integral is localized on the vector multiplet's scalar field $\s=\text{constant}$ and $D=-\frac{\s}{r}$, we have,
\equ{\W(x,θ,\bar{θ})=\frac{1}{\dim R}\Tr_R\P\exp∫dτ\big[\tfrac{-\i}{2}\dot{x}_A^{αβ}θ_α\bar{θ}_β\,D +\dot{θ}·\(\bar{θ}\,\s +θ\bar{θ}^2D\) +|\dot{x}_A|\(\s+θ·\bar{θ}D\)\!\big].
\label{WLN2}}
Even though we do not know $f(τ)$ explicitly, we can evaluate $\braket{\W(x,θ,\bar{θ})}$ formally. Let us denote everything in the exponent by $Θ\s$, with $Θ=\frac{1}{2π}∫dτ\big(1 +\frac{\i}{2}\dot{x}_A^{αβ}θ_α\bar{θ}_β+\dot{θ}·\bar{θ} -θ·\bar{θ} +\dot{θ}·θ\bar{θ}^2\big)$ (also set $r=1$). The path integral reduces to a matrix model in terms of eigenvalues $λ_i$ of $\s$ (we choose ABJM for concreteness, which has two $U(N)$'s as gauge groups and $±k$ as the two Chern-Simons levels):
\equ{\braket{\W(x,θ,\bar{θ})}=\frac{1}{N!N\,Z}∫dλ_i\,d\hat{λ}_i\(e^{-\frac{N}{2α}λ_i^2-\frac{N}{2\hat{α}}\hat{λ}_i^2}\)Δ(λ)^2Δ(\hat{λ})^2{\textstyle\(∑_i e^{Θ λ_i}\)}×Z_{\text{1-loop}}.
}
where $α=-\hat{α}=2π\i \frac{N}{k}$. We refer the reader to \cite{Kapustin:2009kz} for the definitions of various factors in the above result as we are interested in its perturbative limit only. To obtain a perturbative $α$ expansion, we can expand $\braket{\W}$ in $λ$ and compute the vev using the orthogonal polynomials method $\big($note that $\braket{λ^{2k}}=\O(α^k)\big)$:
\equ{\braket{\W(x,θ,\bar{θ})}=1 +\frac{1}{2}Θ^2 α -\left[\frac{1}{6}\(1+\frac{1}{2N^2}\)Θ^2 -\frac{1}{24}\(2+\frac{1}{N^2}\)Θ^4\right]α²+\O(α^3).
}
Rewriting $Θ=1+\frac{1}{2}ϑ$, we get (note $ϑ^3=0$)
\equ{\braket{\W(x,θ,\bar{θ})}=1 +\frac{1}{2}\left[ϑ+\frac{ϑ^2}{4}\right]α -\left[\frac{1}{24}\(5+\frac{1}{N^2}\) +\frac{1}{4}ϑ -\frac{1}{6}\(\frac{1}{2}+\frac{1}{N^2}\)\frac{ϑ^2}{4}\right]α²+\O(α^3).
}
In the above expression, we have removed the bosonic term at $\O(α)$ by multiplying the result by an overall phase $e^{-\frac{1}{2}α}$, which is necessary in matching the perturbative computation~\cite{Kapustin:2009kz}. Note that we do not remove the whole $ϑ$-dependent term at $\O(α)$, since as we will see later there are indeed fermionic contributions at $\O(α)$ in perturbative computation. We will return back to this result in Section \ref{Sc:com}.

\section[\texorpdfstring{Review of $\N=3$ Harmonic Superspace}{Review of N=3 Harmonic Superspace}]{Review of $\bs{\N=3}$ Harmonic Superspace}\label{Sc:Rev}
Now, we turn to $\N=3$ supersymmetry. We collect here the necessary ingredients from three-dimensional $\N=3$ harmonic superspace literature along with a few explicitly worked out details that will be relevant for us in later sections.

\subsection[\texorpdfstring{$\N=3$ Harmonic Superspace}{N=3 Harmonic Superspace}]{$\bs{\N=3}$ Harmonic Superspace}
The `ordinary' $d=3$, $\N=3$ superspace with coordinates $\{x^{αβ},θ^α_{ij}\}$ has the following algebra of superspace derivatives:
\eqsc{\{D_α^{ij},D_β^{kl}\}=\i\big(ε^{ik}ε^{jl}+ε^{il}ε^{jk}\big)∂_{αβ} \nonumber\\
D_α^{ij}=∂_α^{ij}+\i θ^{ij}∂_{αβ}\,.
}
To obtain constrained superfields in the form of $D^{ij}_{\alpha}\Phi=0$, it is useful to consider the case where $D^{ij}_{\alpha}$ is given by a simple partial derivative, indicating the independence of $\Phi$ on certain variables. The obstacle to having a representation of $D^{ij}_{\alpha}$ as a partial derivative is its anti-commutator algebra. This can be overcome by the introduction of $SU(2)/U(1)$ harmonics $u_i^±$. These bosonic variables satisfy
\equ{u^{+i}u^-_i=1,\quad u^{\pm i}u^{\pm}_i=0\,,
}
where the raising and lowering of the $SU(2)$ index $i$ is done by contracting with the invariant tensor $\epsilon^{ij}$. (The contracted $i$ among $u$'s will be suppressed most of the time.) These new variables are to be integrated away using the following rules:
\equ{\tint du \,1=1,\quad \tint du\; u_{(i_1}^{+}\cdots u_{i_n)}^{-}=0.
}
In other words, only the $SU(2)$ invariant polynomial with vanishing $U(1)$ charge survives the integration. The harmonic variables allow us to linearly recombine the $3\times2$ fermionic coordinates into three new $SL(2,\bR)$ doublets $θ^{α,±±}≡u^{i±}u^{j±}θ^α_{ij}$, $θ^{α,0}≡u^{i+}u^{j-}θ^α_{ij}$. The upshot is that doing the same for the covariant derivatives, the supersymmetry algebra now reads, 
\eqsc{\{D_α^{++},D_β^{--}\}=2\i ∂_{αβ}\,,\qquad \{D_α^0,D_β^0\}=-\i ∂_{αβ}\,,\nonumber \\
\{D_α^{±±},D_β^{±±}\}=0\,,\qquad \{D_α^{±±},D_β^0\}=0\,,
}
where one finds that we can $SU(2)$ covariantly isolate a doublet of commuting fermionic derivatives, for example  $D^{++}_{\alpha}$. This implies that we can have a representation for the covariant derivatives where $D^{++}_{\alpha}$ is a simple partial derivative. This is referred to as the ``analytic basis'', and it is given as the following:
\eqsg{∂_{αβ} &→ ∂_{αβ}^A \\
D_α^{ij} &→ \left\{\begin{array}{l}
D_α^{++} = ∂_α^{++} \\
D_α^{--} = ∂_α^{--} +2\i θ^{--β}∂_{αβ}^A \\
D_α^0 = -\half ∂_α^0 +\i θ^{0β}∂_{αβ}^A\,. \\
\end{array}\right.
}
We defined $x_{αβ}^A=x_{αβ}+\i θ_{(α}^{++}θ_{β)}^{--}$. In the analytic basis, we obtain constrained superfields by imposing the `analytic' constraint $D_α^{++}Φ=0$, which now implies that $Φ$ does not depend on $θ_α^{--}$:
\equ{D_α^{++}Φ=0\quad ⇒ \quad Φ≡Φ(x_{αβ}^A,θ_α^{++},θ_α^0,u).
}

The introduction of harmonic variables also introduces R-symmetry covariant derivatives, and are given by\footnote{$D^0$ is strictly speaking not a covariant derivative on $SU(2)/U(1)$. It should be treated as the subgroup generator that defines the $U(1)$ charge for a given operator or field, as in $D^0Φ^{(q)}=qΦ^{(q)}$.}:
\equ{D^{±±}≡∂^{±±}=u_i^{±}\tfrac{∂}{∂u_i^∓}\,,\qquad D^0=[D^{++},D^{--}]\,.
}
These have non-trivial commutator algebra with the fermionic derivatives:\footnote{For completeness, their explicit forms in the analytic basis is given as
\[D^{±±} → D^{±±}=∂^{±±} ±2\iθ^{±±α}θ^{0β}∂_{αβ}^A +2θ^{0α}∂_α^{±±} +θ^{±±α}∂_α^0.\]}
\equ{[D^{±±},D_α^{∓∓}]=2D_α^0\,,\qquad [D^{±±},D_α^0]=D_α^{±±}\,.
}

\subsection{Chern-Simons Matter Theories}
To study gauge theories, we gauge-covariantize the full superspace derivatives $D→\D=D+A$, which define the relevant field strengths:
\eqsc{\{\D_α^{++},\D_β^{--}\}=2\i \D_{αβ}+2ε_{αβ}W^0\,,\qquad \{\D_α^0,\D_β^0\}=-\i \D_{αβ}\,, \\
\{\D_α^{±±},\D_β^{±±}\}=0\,,\qquad \{\D_α^{±±},\D_β^0\}=±ε_{αβ}W^{±±}\,.
}
The covariant derivatives, and the field-strengths, transforms as $\D\rightarrow e^{\tau} \D e^{-\tau}$. Choosing a suitable `gauge-frame' (from $τ→λ$) such that   $A_α^{++}=0$, allows us to define analytic super fields covariantly while maintaining its implication of independence on $θ_α^{--}$: $\D_α^{++}Φ=D_α^{++}Φ=0$. Note that choosing such a gauge generates (new) harmonic connections $A^{±±}$, from which all other connections can be obtained through Bianchi identities. In particular, $A^{++}$ turns out to be the unique analytic ($D^{++}_αA^{++}=0$) prepotential in this formalism. The prepotential transforms under a gauge variation as usual: 
\equ{A^{++'}=e^{\lambda}\D^{++}e^{-\lambda} \quad ⇒ \quad δ_λA^{++}=-\D^{++}λ,}
where $λ$ is an analytic gauge parameter. A convenient gauge is the Wess-Zumino gauge in which the prepotential has the following component expansion\cite{Buchbinder:2009dc}:
\eqsc{\half\D^0_α\D^{--}_βA^{++}{}_|=a_{αβ}\,;\quad \half(\D^0_β)^2\D^{--}_αA^{++}{}_|=2λ_α\,;\quad \half \D^0_α(\D^{--}_β)^2A^{++}{}_|=3χ^{--}_α\,; \nn
\half(\D^{--}_α)^2A^{++}{}_|=3φ^{--}\,;\quad \half(\D^0_α)^2\half(\D^{--}_β)^2A^{++}{}_|=3X^{--}.
\label{AppCexp}}
This is clearly the $\N=3$ vector multiplet with fields $\big(a_μ,λ_α,χ_α^{(ij)},φ^{(ij)},X^{(ij)}\big)$. Of course, $φ^{--}=u^-_iu^-_jφ^{ij}$ and so on.

It is now possible to write every other connection and field strength in terms of the analytic prepotential $A^{++}$. We start with the connections:
\eqsg{D^0=[\D^{++},\D^{--}]\quad &⇒\quad A^{--}(u) =∑_{n=1}^∞(-1)^n∫du_{1,⋯,n} \frac{A^{++}_1⋯A^{++}_n}{(u^+u_1^+)(u_1^+u_2^+)⋯(u_n^+u^+)}\,·\\
2 \D_\alpha^0=[\D^{--},D_\alpha^{++}] \quad &⇒\quad A_\alpha^0 =-\half D_\alpha^{++}A^{--}\,,\\
\D_α^{--}=[\D^{--},\D_α^0] \quad &⇒\quad A_α^{--} =D^{--}A_α^0 -D_α^0A^{--} +[A^{--},A_α^{0}]\,.\\
-i\D_{\alpha\beta}=\{\D^0_{\alpha},\D^0_{\beta}\}\quad &⇒\quad A_{αβ} =2\i D_{(α}^0A_{β)}^0 =-\i D_{(α}^0D_{β)}^{++}A^{--}\,.\label{Amu}
}
Then the covariant field strengths can be derived from the connections as follows: 
\eqsc{D_α^{++}W^{++}=\D^{++}W^{++}=0 \quad ⇒\quad W^{++} \text{ is analytic.} \nonumber\\
W^{++}=\half D_α^{++}A^{0α} =-\tfrac{1}{4}(D_α^{++})^2A^{--}\,. \label{Ws}\\
W^0=\half \D^{--}W^{++}\quad\text{ and }\quad W^{--}=\D^{--}W^0\,.\nonumber
}

The $\N=3$ matter multiplet consists of two complex scalars $f^i$ transforming as a doublet under $SU(2)$ and their fermionic partners $ψ_α^i$, which are encoded in the following hypermultiplet superfield:
\eqsc{q^+{}_|=f^+\,;\quad \D^{--}_αq^+{}_|=ψ^-_α\,;\quad \half\D^0_α\D^{--}_βq^+{}_|=-\i ∂_{αβ}f^-\,, \nn
\bar{q}^+{}_|=-\bar{f}^+\,;\quad \D^{--}_α\bar{q}^+{}_|=\bar{ψ}^-_α\,;\quad \half\D^0_α\D^{--}_β\bar{q}^+{}_|=\i ∂_{αβ}\bar{f}^-\,.
}
where $f^{\pm}\equiv u_{i}^\pm f^{i}$, $\bar{f}^{\pm}\equiv u_{i}^\pm\bar{f}^{i}$, and similarly for the fermions.

For the ABJM theory, we have two sets of $q^{+a}$, with $a=1,2$. In this representation, the $SO(6)$ R-symmetry is broken: $SO(6)\,\rightarrow\,SU(2)_{\text{R}}\times SU(2)_{\text{ext}}$, and the ABJ(M) action for $U_L(N)×U_R(M)$ theory:
\eqsc{\S=\S_{CS}[A_L^{++}]-\S_{CS}[A_R^{++}]+\Tr ∫dζ^{(-4)}\bar{q}^+_a\D^{++}q^{+a}, \\
\S_{CS}[A^{++}]=\frac{\i k}{4π}\Tr ∑_{n=2}^∞\frac{(-1)^n}{n}∫d^3x\,d^6θ\,du_{1,⋯,n}\frac{A^{++}_1⋯A^{++}_n}{(u_1^+u_2^+)⋯(u_n^+u_1^+)}\,, \\
(\D^{++}q^{+a})_A^{\bar{B}}=D^{++}(q^{+a})_A^{\bar{B}} +(A_L^{++})_A^B(q^{+a})_B^{\bar{B}} -(q^{+a})_A^{\bar{A}}(A_R^{++})_{\bar{A}}^{\bar{B}}\,,
}
where $A∈U(N),\,\bar{A}∈U(M)$, and $\bar{q}^{+a}$ have `opposite' gauge charges under the two gauge groups. From the action, one finds the following equations of motion:
\equ{\frac{\delta \S}{\D\bar{q}^+_a}=∇^{++}q^{+a}=0\,,\qquad \frac{\delta \S}{\delta A^{++}}=W^{++} +\frac{4π\i}{k}\bar{q}^+_aq^{+a} =0\,,
\label{EoMs}}
with proper ordering of $\bar{q}q$ to match the gauge indices of $W^{++}_{L,R}$. The latter equation of motion implies that scalars from the vector multiplet get equated to bi-scalars of the matter multiplet. One such relation will be relevant for later use:
\equ{W^0=\half D^{--}W^{++} ⇒ φ^0 =-\tfrac{2π\i}{k}u_i^-\tfrac{∂}{∂u_i^+}\((-u_j^+\bar{f}^j_a)(u_k^+f^{ka})\) =\tfrac{2π\i}{k} (u_j^-u_k^+ +u_j^+u_k^-)\bar{f}^j_a f^{ka}.
\label{Wqq0}}
This crucial relation is responsible for generating the well-known sextic potential involving $f$'s once $φ$'s are integrated out from the ABJM action.

The CS theories coupled to matter can be quantized directly in superspace \cite{Buchbinder:2009dc} and the resulting propagators read:
\eqs{\braket{\bar{q}^+_1q^+_2} &=\frac{1}{2π\i}\frac{u_1^+u_2^+}{\sqrt{2ρ^2}}\,, \\
\braket{A^{++}_1A^{++}_2} &=\frac{\i}{2π}\frac{1}{\sqrt{2ρ^2}}δ^2(θ^{++}_{12})δ^{(-2,2)}(u_1,u_2)\,,
}
where
\eqs{ρ^{αβ}&=x_{A\,1}^{αβ} -x_{A\,2}^{αβ} -2\i θ^{0(α}_1 θ^{0β)}_2 -\tfrac{2\i}{u^+_1u^+_2}\Big[(u^-_1u^-_2)θ^{++(α}_1 θ^{++β)}_2 -(u^-_1u^+_2)θ^{++(α}_1 θ^{0β)}_2 \nn
&\qquad-(u^+_1u^-_2)θ^{0(α}_1 θ^{++β)}_2 +(u^-_1u^+_2)θ^{++(α}_1 θ^{0β)}_1 +(u^+_1u^-_2)θ^{0(α}_2 θ^{++β)}_2\Big]\,.
}
The $ρ^{αβ}$ has quite a complicated expression but in the presence of $δ^2(θ^{++}_{12})δ(u_1,u_2)$, it simplifies in the vector propagator to the following:
\equ{ρ^{αβ}=\big(x_A^{αβ}\big)_{12} -2\i θ^{0(α}_1 θ^{0β)}_2 \quad ⇒ \quad ρ^2=-2|x^A_{12}|^2-4\i
x^A_{12}·θ^0_1θ^0_2 +4{θ^0_1}^2{θ^0_2}^2\,.
}
The vertices are easily read from the relevant actions.

\section{Super-Wilson Loop}\label{Sc:SWL}
There are two main types of Wilson loop operators that can be considered for $d=3$ Chern-Simons theories \cite{Gaiotto:2007qi, Kapustin:2009kz, Drukker:2009hy, Ouyang:2015bmy}: GY-type ($\frac{1}{\N}$-BPS for $\N=2,3,4,6$) and DT-type (still $\frac{1}{\N}$-BPS for $\N=2,3$ but $\half$-BPS for $\N=4,6$). We will focus only on the former case here. The $\frac{1}{3}$-BPS Wilson loop is usually written for $\N=3$ CS theory as follows:
\equ{\W_{\sfrac{1}{3}}(x)=\tfrac{1}{\dim R}\Tr_R \P \exp{\textstyle∫}dτ\left[\tfrac{-\i}{2}\dot{x}^{αβ} a_{αβ} +\half\dot{y}_{ij} φ^{ij}\right],
\label{bosWL}}
where $y_{ij}=y_{ji}$ are 3 $SU(2)$ `coordinates'. For this operator to locally preserve any supersymmetry, the susy parameter $ε^{ij}_α$ needs to be a solution of
\equ{\dot{x}^{αβ}\epsilon_{\beta}^{ij} +\dot{y}^i{}_k\epsilon^{\alpha,kj}=0\,,
\label{N3bpsEq}}
provided that $|\dot{x}|=|\dot{y}|$. To incorporate the condition on $|\dot{y}|$, we can rewrite the scalar term in WL as $∫dτ|\dot{x}|(u^+_iu^-_j)φ^{(ij)}$ using the harmonic coordinates on $SU(2)$.

Now, we are ready to write down the most general supersymmetrized expression for a Wilson loop (such that \eqref{bosWL} is its bosonic component):
\eqst{\W(x,θ^{±±},θ^0)=\frac{1}{\dim R}\Tr_R \P \exp ∫ dτ\bigg[\frac{-1}{4}\dot{x}^{A,αβ} A_{αβ} +\dot{θ}^{++α} A^{--}_α +\dot{θ}^{0α} A^0_α +{\textstyle ∑_±}u^{±i}\dot{u}^±_i A^{∓∓} \\
+\frac{1}{2}|\dot{x}^A|W^0\bigg].
\label{WL13t}}
The usual BPS condition on the bosonic WL $(ε^{(ij)γ}Q^{(ij)}_γ\W_{\sfrac{1}{3}}(x)=0)$, which results in \eqref{N3bpsEq}, translates to $ε^{(ij)γ}D^{(ij)}_γ$ $\W_{\sfrac{1}{3}}=0$ (along with $\dot{x}→\dot{x}^A$) in superspace for obvious reasons (see \cite{Beisert:2015jxa} for an explicit proof). Let us see what that implies for \eqref{WL13t}: 
\[\begin{aligned}
ε^{(ij)γ}D^{(ij)}_γ\W(x,θ^{++},θ^0) \propto ∫dτ \Big[&\tfrac{-1}{4}\dot{x}^{A,αβ} ε^{(ij)γ}\F^{(ij)}_{γ,αβ} +\dot{θ}^{++α}ε^{(ij)γ}\F^{(ij),--}_{γ,α} +\dot{θ}^{0α}ε^{(ij)γ}\F^{(ij),0}_{γ,α} \\
&+\underbrace{0}_{\F^{(ij),∓∓}_γ≡0} +\tfrac{1}{2}|\dot{x}^A|ε^{(ij)γ}\D^{(ij)}_γW^0\Big], 
\end{aligned}\]
where we use $\F_{A,B}$ to represent the field strength arising from the (anti-)commutator of $\{\D_A, \D_B]$.  As we did for the case of $\N=2$ WL, we have ignored here terms that look like field-dependent gauge transformations. Since we know that only $\F^{0,0}_{γ,α}=\F^{±±,±±}_{γ,α}=0$, we can have only one of the $\dot{θ}$ terms above in the Wilson loop. This means either $ε^{++}$ or $ε^0$ can be the only unbroken susy. However, choosing $ε^{0}$, we find that $\F^{0}_{γ,αβ}$ \cite{Buchbinder:2009dc} contains not only the $\D^0_αW^0$ term but also $\D^{++}_αW^{--}$ so the above variation cannot vanish. Thus we are left with $ε^{++}$ and the remaining couple of terms do vanish in this case because
\equ{\F^{--}_{γ,αβ} = -\i\(ε_{γα}\D^{--}_βW^0 +ε_{γβ}\D^{--}_αW^0\),
}
which implies
\eqst{-\half\dot{x}^{A,αβ} ε^{++γ}\F^{--}_{γ,αβ} +|\dot{x}^A|ε^{++γ}\D^{--}_γW^0 = ε^{++γ}\(\i\dot{x}^{A,αβ}ε_{γα}\D^{--}_βW^0 +|\dot{x}^A|\D^{--}_γW^0\) \\
= \i ε^{++}_α\(\dot{x}^{A,αβ} -\i|\dot{x}^A|ε^{αβ}\)\D^{--}_βW^0 =0 \,.
}
This expression vanishes (for arbitrary $W^0$) in a way similar to the $\N=2$ case, and we preserve half of the complex spinor $ε^{++γ}$. Thus, the final result for the supersymmetric generalization of the $\frac{1}{3}$-BPS Wilson loop is:
\equ{\W_{\sfrac{1}{3}}=\frac{1}{\dim R}\Tr_R \P \exp∫dτ\Big[\tfrac{-1}{4}\dot{x}^{A,αβ} A_{αβ} +\dot{θ}^{++α} A^{--}_α +{\textstyle ∑_±}u^{±i}\dot{u}^±_i A^{∓∓} +\tfrac{1}{2}|\dot{x}^A|W^0\Big].
\label{WL13}
}
To compare with the usual bosonic WL operator, we write the above in component fields
\equ{\W_{\sfrac{1}{3}}\sim\Tr_R \P \exp∫dτ\Big[\tfrac{-\i}{2}\dot{x}^{αβ} a_{αβ} +\dot{θ}^{++α} θ^{--}_αφ^0 +\tfrac{1}{2}|\dot{x}^A|(φ^0+θ^0·χ^0) +\O(θ^2)\Big].
\label{WL13cfs}
}
The difference starts at terms of order $θ$ containing fermionic fields ($χ_α^{ij}$) and at $θ^2$ order with bosonic fields ($φ^{ij}$). Higher-order terms will contain $λ_α, X^{ij}$ fields too.

With this construction, we can readily give the supersymmetrized generalization of the $\frac{1}{6}$-BPS WL operator for $\N=6$ ABJM theory in $\N=3$ harmonic superspace:
\eqst{\W_{\sfrac{1}{6}}=\frac{1}{\dim R}\Tr_R \P \exp∫dτ\left[\left(\tfrac{-1}{4}\dot{x}^{A,αβ} A_{αβ} +\dot{θ}^{++α} A^{--}_α +{\textstyle ∑_±}u^{±i}\dot{u}^±_i A^{∓∓} +\tfrac{1}{2}|\dot{x}^A|W^0\right)_L\right. \\
\left.+\Big(L→R\Big)\right].
\label{WL16}
}
This operator reduces to the canonical bosonic operator in ABJM theory with the matter coupling term $M_I^JC_J\bar{C}^I$ where $M_I^J=\text{diag}(-1,-1,1,1)$ (up to the factor $\frac{2π}{k}$) if the $u$-matrix further satisfies $u^+_1=(u^-_2)^{-1}=u(τ)$. To show this, we need to use the equation of motion for $A^{++}$ \eqref{EoMs} and \eqref{Wqq0} along with a change of notation from $f^i → C_I$ as discussed in \cite{Buchbinder:2008vi}. (Without the constraint on $u$, this operator has more content due to $W^0$ containing not only $φ^{12}≡M_I^JC_J\bar{C}^I$ but also $φ^{11}$ and $φ^{22}$.)

\section{Computation}\label{Sc:Com}
In this section, we will compute the `one-loop' vacuum expectation value of the Wilson loop $\W_{\sfrac{1}{3}}$. The constraint on $u$ will also be imposed so the operator and the expected vev slightly simplify (with $R$ being the fundamental representation of $U(N)$ gauge group):
\eqs{\W_{\sfrac{1}{3}} &=\tfrac{1}{N}\Tr_R \P \exp\tint dτ\big[\,\tfrac{-1}{4}\dot{x}^{A,αβ} A_{αβ} +\dot{θ}^{++α} A^{--}_α +\tfrac{1}{2}|\dot{x}^A|W^0\,\big] \label{WL13s} \\
\braket{\W_{\sfrac{1}{3}}} &= 1 +\frac{1}{2N}∫dτ_1dτ_2\braket{\Big(\tfrac{-1}{4}\dot{x}^{A}·A +\dot{θ}^{++}· A^{--} +\tfrac{1}{2}|\dot{x}^A|W^0\Big)_1\Big(⋯\Big)_2} +⋯\,.
}

An important subtlety that occurs repeatedly in the computation is when $D^{++}_{\alpha}(u)=u^+_{i}u^+_{j}D^{ij}_α$ acts on an analytic superfield which depends on another harmonic variable, say $(\theta_{\alpha}^{++},\theta_{\alpha}^0,u')$. The result is not the naive zero since using $(u^+u^-)=1$ and repeated Schouten identities, one can rewrite:
\eqs{D^{++}_{\alpha}(u) &=(u'^+u'^-)^2u^+_{i}u^+_{j}D^{ij}_\alpha \nonumber\\
&=\left[ (u^+u'^-)^2D_{α}^{++}(u') +(u^+u'^+)^2D_{α}^{--}(u') -2(u^+u'^-)(u^+u'^+)D^0_{α}(u')\right],
}
where we have converted the harmonic dependence of the derivative from $u$ to $u'$. Note that the charges match on both sides separately for $u$'s and $u'$'s. Thus for any analytic superfield $\Phi(\theta_{\alpha}^{++},\theta_{\alpha}^0,u')$, we have:
\equ{D^{++}_{\alpha}(u)\Phi(u')= (u^+u'^+)^2\left[D_{α}^{--}\Phi\right](u') -2(u^+u'^-)(u^+u'^+)\left[D_{α}^0\Phi\right](u') \,.
\label{D1D22}}
Similar manipulations lead to the following list of identities:
\eqs{\(D^{++}(u)\)^2\Phi(u') &=(u^+u'^+)^2\left((u^+u'^+)^2 \left[(D_{α}^{--})^2\Phi\right](u') -4(u^+u'^-)(u^+u'^+)\left[D^{--\alpha}D^0_{\alpha}\Phi\right](u') \right. \nn
&\quad \left.+4(u^+u'^-)^2 \left[(D^0_{α})^2\Phi\right](u')\right) \nn
D^0_{\alpha}(u)\Phi(u') &=(u^+u'^+)(u^-u'^+)\left[D^{--}_{\alpha}\Phi\right](u')+\(1 -2(u^+u'^+)(u^-u'^-)\)\left[D^0_{\alpha}\Phi\right] (u')\nn
D_{α}^{--}(u)Φ(u') &=(u^-u'^+)^2\left[D_{α}^{--}Φ\right](u') -2(u^-u'^+)(u^-u'^-)\left[D_{α}^0Φ\right](u')\,.
\label{D2D23}}
For the sake of convenience, we list generating expressions for component expansions of some connections and field strengths below $\big($that is, keeping only a single $A^{++}$ in \eqref{Amu} and \eqref{Ws}$\big)$:
\eqs{A_{αβ}(u) &=-\i \big[D^0_{(α}D^{++}_{β)}A^{--}\big] (u)=\i\tint du' \big[D^0_{(α}D^{--}_{β)}A^{++}\big](u') \nn
A^{--}_α(u) &=-\half\big[(D^{--}D^{++}_{α}+2D^0_{α})A^{--}\big](u) \nn
&=-\tint du'\Big[\tfrac{u^-u'^+}{u^+u'^+}\left[D^{--}_{α} A^{++}\right](u')+2\tfrac{u^-u'^-}{u^+u'^+}\left[D^0_{α}A^{++}\right](u')\Big] \label{GenCexp}\\
W^{++}(u) &=-\tfrac{1}{4}\big[(D^{++}_{α})^2A^{--}\big](u) \nn
&=-\tfrac{1}{4}\tint du'\Big[(u^+u'^+)^2\left[(D^{--}_{α})^2A^{++}\right](u') -4(u^+u'^-)(u^+u'^+)\left[D^{--}·D^0A^{++}\right](u')\Big]\,.\nonumber
}
Note that all the fields depend on the same $θ$-coordinate. The components can be obtained from the above expressions by using \eqref{AppCexp} and performing not only the $D$-algebra but some harmonic algebra too. The simplest component to obtain is the vector: $A_{αβ}{}_| = 2\i a_{αβ}$. To get the scalars, we need to perform slightly more involved algebra:
\eqs{W^{++}{}_|&=-\tfrac{1}{4}\tint du'(u^+u'^+)^2\big[(D^{--}_α)^2A^{++}\big](u'){}_| =3\tint du'(u^+u'^+)^2φ^{--}(u') =u^+_j u^+_k φ^{(jk)}=φ^{++}\,; \nn
W^0{}_|&=\tfrac{1}{2}D^{--}W^{++}{}_| =\tfrac{1}{2}\big(u^-_{j}u^+_{k} +u^+_{j}u^-_{k}\big)φ^{(jk)}=φ^0\,; \nn
W^{--}{}_|&=D^{--}W^{0}{}_| =φ^{--}.
}
Other components can be similarly obtained, which we leave as an exercise and refer the reader to \cite{Galperin:2001uw} for useful identities involving harmonic variables.

Now we turn to evaluating various contributions to $\braket{\W_{\sfrac{1}{3}}}$. First let us consider the contribution from the vector connection. In general, we have from \eqref{Amu}:
\eqst{\dot{x}_{A,1}^{αβ}\dot{x}_{A,2}^{γδ}\braket{A_{1,αβ}A_{2,γδ}} = -\dot{x}_{A,1}^{αβ}\dot{x}_{A,2}^{γδ}\langle D^{0}_{1\alpha}D_{1\beta}^{++}A_{1}^{--}D^{0}_{2\gamma}D_{2\delta}^{++}A_2^{--}\rangle \\
= -\dot{x}_{A,1}^{αβ}\dot{x}_{A,2}^{γδ}\left\langle ∫du\frac{D_{1\alpha}^0(u_1)D_{1\beta}^{++}(u_1)A_1^{++}(u)}{(u^+_1u^+)^2} ∫dv\frac{D_{2\gamma}^0(u_2)D_{2\delta}^{++}(u_2)A_2^{++}(v)}{(u^+_2v^+)^2} \right\rangle +⋯
}
Using \eqref{GenCexp}, we find 
\eqs{\dot{x}_{A,1}^{αβ}\dot{x}_{A,2}^{γδ}\braket{A_{1,αβ}A_{2,γδ}}^{(1)} &=- \dot{x}_{A,1}^{αβ}\dot{x}_{A,2}^{γδ}\bcontraction{∫du\big[D_{1α}^0D_{1β}^{--}}{A_1^{++}}{\big](u)∫dv\big[D_{2γ}^0D_{2δ}^{--}}{A_2^{++}}∫du\big[D_{1α}^0D_{1β}^{--}A_1^{++}\big](u)∫dv\big[D_{2γ}^0D_{2δ}^{--}A_2^{++}\big](v) \nn
&= \dot{x}_{A,1}^{αβ}\dot{x}_{A,2}^{γδ}∫du\,D_{1α}^0D_{2γ}^0D_{1β}^{--}D_{2δ}^{--}\frac{\i}{2π\sqrt{2ρ^2}}δ^2(θ^{++}_{12}) \nn
&=-\dot{x}_{A,1}^{αβ}\dot{x}_{A,2}^{γδ}\frac{ε_{βδ}\big(\i x^A_{12} -θ^0_1θ^0_2\big)_{αγ}}{2π\big({x^A_{12}}^2\big)^{\sfrac{3}{2}}} = \dot{x}_{A,1}^{αβ}\dot{x}_{A,2}^{γδ}\frac{ε_{βδ} θ^0_{1α}θ^0_{2γ}}{2π|x^A_{12}|^3}\,·
\label{VV}}
We used here $D_α^{--}D_β^{--}=\half ε_{αβ}{D^{--}}^2$, ${D^{--}}^2δ^2(θ^{++})=4$, $\dot{x}_{A,1}^{αβ}\dot{x}_{A,2}^{γδ}ε_{βδ}x^A_{12,αγ}\sim ε_{mnp}\dot{x}_{A,1}^m\dot{x}_{A,2}^n x_{A,12}^p→0$, and expanded $\frac{1}{ρ^2}$ in powers of $θ^0$'s. The next term (quadratic in $A^{++}$) in the expansion of $A^{--}$ also contributes:
\equ{\dot{x}_{A,1}^{αβ}\dot{x}_{A,2}^{γδ}\braket{A_{1,αβ}A_{2,γδ}}^{(2)}= \frac{\dot{x}_{A,1}·\dot{x}_{A,2}\big(|x^A_{12}|^2 -\i x^A_{12}·θ_1^0θ_2^0 +\frac{1}{4}{θ_1^0}^2{θ_2^0}^2\big)^2}{4π^2|x^A_{12}|^6}\,·
\label{AvAv2}}
Let us sketch how we got this result. We require that all $δ^2(θ^{++}_{12})$ be cancelled so higher orders of $A^{++}$ cannot contribute as there are not enough $D^{--}_α$ derivatives in $\braket{A_{1,αβ}A_{2,γδ}}$ to cancel more than two such $δ$-functions. After expanding $D^0_{1α}D^{++}_{1β}$ using the identities given above, doing two harmonic integrals using the harmonic $δ$-functions in the two propagators and then hitting the two $δ^2(θ^{++}_{12})$ with correct $D^{--}$'s, we are left with:
\eqsn{\braket{A_{1,αβ}A_{2,γδ}}^{(2)} &\sim∫dvdw\,\frac{\begin{array}{c}
\(-(u_1^+v^+)^2(w^+u_1^+)(w^+u_1^-) +(u_1^+w^+)^2(v^+u_1^+)(v^+u_1^-)\)× \\
\(-(u_2^+v^+)^2(w^+u_2^+)(w^+u_2^-)ε_{βδ}ε_{αγ} +(v^+u_2^+)(v^+u_2^-)(w^+u_2^+)^2ε_{βγ}ε_{αδ}\)
\end{array}}{(u_1^+v^+)(w^+u_1^+)(v^+w^+)^2(u_2^+v^+)(w^+u_2^+)\sqrt{2ρ^2}\sqrt{2ρ^2}} \nn
&\sim∫dvdw\,\frac{\begin{array}{c}
-(u_1^+v^+)(w^+u_1^-)\((u_2^+v^+)(w^+u_2^-)ε_{βδ}ε_{αγ} +(v^+u_2^-)(w^+u_2^+)ε_{βγ}ε_{αδ}\) \\
-(u_1^+w^+)(u_1^-v^+)\((u_2^+v^+)(w^+u_2^-)ε_{αδ}ε_{βγ} +(v^+u_2^-)(w^+u_2^+)ε_{αγ}ε_{βδ}\)
\end{array}}{(v^+w^+)^2\(2ρ^2\)} \nn
&\sim∫dvdw\,\frac{\begin{array}{c}
-\((u_1^+w^+)(u_1^-v^+)+(v^+w^+)\)(v^+w^+)ε_{βγ}ε_{αδ} \\
+(u_1^+w^+)(u_1^-v^+)(v^+w^+)ε_{αγ}ε_{βδ}
\end{array}}{(v^+w^+)^2\(2ρ^2\)} \nn
&\sim\frac{ε_{αγ}ε_{βδ}}{ρ^2} =\frac{ε_{αγ}ε_{βδ}\big(|x^A_{12}|^2 -\i x^A_{12}·θ_1^0θ_2^0 +\frac{1}{4}{θ_1^0}^2{θ_2^0}^2\big)^2}{|x^A_{12}|^6}\,·
}
Keeping track of various signs and numerical factors above, we get \eqref{AvAv2}.

Let us now evaluate the second contribution to $\braket{\W}$ due to the `charged' fermionic connection. Using the fact that we need enough $D^{--}_α$ to get relevant terms, we ignore terms with $D^0_α$ in the expansion of $A^{--}$ in \eqref{GenCexp}:
\eqs{\dot{θ}^{++}_{1α}\dot{θ}^{++}_{2β}\braket{A_1^{--α}A_2^{--β}} &=\dot{θ}^{++}_{1α}\dot{θ}^{++}_{2β}∫du\(\tfrac{u^+u_1^-}{u^+u_1^+}\)D_1^{--α}\(\tfrac{u^+u_2^-}{u^+u_2^+}\)D_2^{--β}\frac{\i δ^2(θ^{++}_{12})}{2π\sqrt{2ρ^2}} \nn
&=\dot{θ}^{++}_{1}·\dot{θ}^{++}_{2}\(\frac{u_1^-u_2^-}{u_1^+u_2^+}\)\frac{|x^A_{12}|^2 -\i x^A_{12}·θ_1^0θ_2^0 +\frac{1}{4}{θ_1^0}^2{θ_2^0}^2}{2π |x^A_{12}|^3}\,·
}
The $u$-factor in parentheses might look divergent upon imposing the constraint on $u$-matrix discussed in the previous section but using an explicit parameterization, one can show that it instead limits to unity up to a $U(1)$ `charge factor'. We will, however, leave this factor as it is to account for the correct $U(1)$ charges along with an understanding that there is no non-trivial $u$-dependence.

The third contribution to $\braket{\W}$ due to mixed contraction of the two connections vanishes:
\eqs{\braket{A_{1,αβ}A_2^{--γ}} &= -\i∫du\,D_{1(α}^0D_{1β)}^{--}\(\tfrac{u^+u_2^-}{u^+u_2^+}\)D_2^{--γ}\frac{\i δ^2(θ^{++}_{12})}{2π\sqrt{2ρ^2}} \nn
&=\frac{-\big(x^A_{12,\s(α}θ_2^{0\s} +\frac{\i}{2}θ_{1(α}^0 {θ_2^0}^2\big)δ_{β)}^γ}{2π|x^A_{12}|^3}∫du\(\tfrac{u^+u_2^-}{u^+u_2^+}\) =0\,.
}

The fourth contribution to $\braket{\W}$ due to the scalar field strength is
\eqs{|\dot{x}_{A,1}||\dot{x}_{A,2}|\braket{W_1^0 W_2^0}^{(1)} &=\tfrac{1}{64} |\dot{x}_{A,1}||\dot{x}_{A,2}|\bcontraction{\D_1^{--}{D_{1α}^{++}}^2}{A_1^{--}}{\D_2^{--}{D_{2β}^{++}}^2}{A_2^{--}}\D_1^{--}{D_{1α}^{++}}^2A_1^{--}\D_2^{--}{D_{2β}^{++}}^2A_2^{--} \nn
&=\frac{|\dot{x}_{A,1}||\dot{x}_{A,2}|\big({θ^0_1}^2 +θ^0_1·θ^0_2 +{θ^0_2}^2\big)}{2π|x^A_{12}|^3}\(1 -2(u_1^+u_2^+)(u_1^-u_2^-)\).
}
This is a contribution from the linear term in $A^{--}$ and is straightforward to compute. Like $\braket{A_{αβ}A_{γδ}}$, we get a second contribution from the contraction of quadratic terms in $A^{--}$ here too: 
\equ{|\dot{x}_{A,1}||\dot{x}_{A,2}|\braket{W_1^0 W_2^0}^{(2)} = -\frac{|\dot{x}_{A,1}||\dot{x}_{A,2}|\big(|x^A_{12}|^2 -\i x^A_{12}·θ_1^0θ_2^0 +\frac{1}{4}{θ_1^0}^2{θ_2^0}^2\big)^2}{4π^2|x^A_{12}|^6}\(1 -2(u_1^+u_2^+)(u_1^-u_2^-)\).
}
This computation proceeds very similarly to the case of the vector connection but there are more terms; we sketch them below (again, various signs and numerical factors need to be tracked):
\eqsn{\braket{W^0_1W^0_2}^{(2)}\sim&∫dv_{1,2}\frac{\D_1^{--}{D_{1α}^{++}}^2[A^{++}(v_1)A^{++}(v_2)]}{(u_1^+v_1^+)(v_1^+v_2^+)(v_2^+u_1^+)}∫dw_{1,2}\frac{\D_2^{--}{D_{2β}^{++}}^2[A^{++}(w_1)A^{++}(w_2)]}{(u_2^+w_1^+)(w_1^+w_2^+)(w_2^+u_2^+)} \\
\sim& D_1^{--}D_2^{--}∫dv_{1,2}\frac{{D_{α}^{--}(v_1)}^2{D_{β}^{--}(v_2)}^2\left[\frac{(u_1^+v_1^+)^4⋯}{\sqrt{2ρ^2}}δ^2(θ^{++}_{12})\frac{(u_2^+v_2^+)^4⋯}{\sqrt{2ρ^2}}δ^2(θ^{++}_{12})\right]}{(u_1^+v_1^+)(v_1^+v_2^+)(v_2^+u_1^+)(u_2^+v_1^+)(v_1^+v_2^+)(v_2^+u_2^+)}\\
\sim& D_1^{--}D_2^{--}∫dv_{1,2}\frac{(u_1^+v_1^+)^2(u_1^+v_2^+)^2 (u_2^+v_1^+)^2(u_2^+v_2^+)^2}{(u_1^+v_1^+)(v_1^+v_2^+)(v_2^+u_1^+)(u_2^+v_1^+)(v_1^+v_2^+)(v_2^+u_2^+)ρ^2} \\
\sim& D_1^{--}D_2^{--}∫dv_{1,2}\left[\frac{(u_1^+v_1^+)(u_1^+v_2^+)(u_2^+v_1^+)(u_2^+v_2^+)}{(v_1^+v_2^+)^2}\right]\frac{1}{ρ^2} \\
\sim& \(2-4(u_1^+u_2^+)(u_1^-u_2^-)\) \tfrac{1}{ρ^2}\,·
}
Similarly, we can compute two more mixed contractions between the two connections and $W^0$, but only one is non vanishing:
\equ{\braket{A_{1α}^{--}W_2^0} = -\frac{\big(\i x^A_{12}·θ^0_1 -\frac{1}{2}θ_2^0\,{θ_2^0}^2\big)_α}{2π|x^A_{12}|^3}\(u^-_1u^+_2\)\(u^-_1u^-_2\).
}

One more contribution to $\braket{\W}$ needs to be considered (at the order being studied) and this one includes a 3-point vertex insertion:
\eqs{&\braket{\tint dτ_1\dot{x}_{A,1}·A_1\tint dτ_2\dot{x}_{A,2}·A_2\tint dτ_3\dot{x}_{A,3}·A_3} \nn
=& -\i\tint dτ_{1,2,3}\,\dot{x}^{αβ}_{A,1}\dot{x}^{γδ}_{A,2}\dot{x}^{κλ}_{A,3}\tint dv_{1,2,3}(D^0_αD^{--}_βA^{++})_1(D^0_γD^{--}_δA^{++})_2(D^0_κD^{--}_λA^{++})_3 \nn
\sim& \tint d^3τ⋯\tint d^3v(D^0_αD^{--}_β)_1(D^0_γD^{--}_δ)_2(D^0_κD^{--}_λ)_3\tint d^3x_0d^6θ_0\tfrac{dw_{1,2,3}}{(w^+_1w^+_2)(w^+_2w^+_3)(w^+_3w^+_1)}{\textstyle ∏_{i=1}^3}\tfrac{δ^2(θ^{++}_{0i})δ(v_i,w_i)}{\sqrt{2ρ^2_{0i}}} \nn
\sim& \tint d^3τ⋯\tint d^3v\,(v^+_1v^+_2)(v^+_2v^+_3)(v^+_3v^+_1)\tint d^3x_0(D^0_αD^{--}_β)_{1,2,3}{\textstyle ∏_{i=1}^3}\tfrac{|x_{0i}|^2+\i(x_{0i})·(θ^{++}_iθ^{--}_i) -\frac{1}{4}{θ^{++}_i}^2{θ^{--}_i}^2}{|x_{0i}|^3} \nn
\sim& \tint d^3τ\,\dot{x}^{αβ}_{A,1}\dot{x}^{γδ}_{A,2}\dot{x}^{κλ}_{A,3}\tint d^3v\,(v^+_1v^+_2)(v^+_2v^+_3)(v^+_3v^+_1)(v^-_1v^+_2)(v^-_2v^+_3)(v^-_3v^+_1)(v^-_1v^-_2)(v^-_2v^-_3)(v^-_3v^-_1) \nn
&×\int d^3x_0\(\frac{(x_{01})_{βγ}(x_{02})_{δκ}(x_{03})_{λα}}{|x_{01}|^3|x_{02}|^3|x_{03}|^3} -\frac{\i(x_{01})_{βγ}(x_{02})_{δκ}θ^{++}_{3,λ}θ^{--}_{3,α}}{|x_{01}|^3|x_{02}|^3|x_{03}|^3}-⋯+\frac{\i θ^{++}_{1,β}θ^{--}_{1,γ}θ^{++}_{2,δ}θ^{--}_{2,κ}θ^{++}_{3,λ}θ^{--}_{3,α}}{|x_{01}|^3|x_{02}|^3|x_{03}|^3}\).
}
The second to last line is obtained after performing the $∫d^6θ_0$ in the previous line with the help of three $δ^2(θ^{++}_{0i})$'s, cancelling the divergent harmonic denominator. The last line then follows by converting $D^0_i→D^{++}_{i+1}$ in cyclic order and acting on the numerator, thus picking out eight terms. The $∫d^3v$ integral produces only a numerical factor. Note that the first term in the integral $∫d^3x_0$ is the only `bosonic' piece given by the well-known integral (6.12) of \cite{Rey:2008bh}.

Finally, collecting all the results at `one-loop' order (we suppress $u$-dependent factors from $\braket{W^0W^0}$ to keep the expression below manageable), we have
\eqs{\braket{\W_{\sfrac{1}{3}}(x,θ^{±±},θ^{0})} =1 &+\frac{1}{2}\frac{4π}{\i k}\frac{N}{2}∫dτ_1dτ_2 \left\{\frac{\dot{x}_{A,1}^{αβ}\dot{x}_{A,2}^{γδ}ε_{βδ}θ^0_{1α}θ^0_{2γ} +|\dot{x}_{A,1}||\dot{x}_{A,2}|\big({θ^0_1}^2 +θ^0_1·θ^0_2 +{θ^0_2}^2\big)}{2π|x^A_{12}|^3} \right. \nn
&\;\; +\left. \dot{θ}^{++}_{1}·\dot{θ}^{++}_{2}\(\frac{u_1^-u_2^-}{u_1^+u_2^+}\)\frac{|x^A_{12}|^2 -\i x^A_{12}·θ_1^0θ_2^0 +\frac{1}{4}{θ_1^0}^2{θ_2^0}^2}{2π |x^A_{12}|^3} \right. \nn
&\;\; -\left. \frac{\dot{θ}^{++}_{1}·\big(\i x^A_{12}·θ^0_2 -\frac{1}{2}θ_1^0\,{θ_2^0}^2\big)|\dot{x}_{A,2}| +|\dot{x}_{A,1}|\dot{θ}^{++}_{2}·\big(\i x^A_{12}·θ^0_1 -\frac{1}{2}{θ_1^0}^2θ_2^0\big)}{2π|x^A_{12}|^3\(u^-_1u^+_2\)^{-1}\(u^-_1u^-_2\)^{-1}} \right\} \nn
&-\frac{1}{2}\frac{16π^2}{k^2}\frac{N^2}{2}∫dτ_1dτ_2\left\{\frac{\dot{x}_{A,1}·\dot{x}_{A,2} -|\dot{x}_{A,1}||\dot{x}_{A,2}|}{4π^2|x^A_{12}|^2}\Bigg(1-\frac{2\i x^A_{12}·θ_1^0θ_2^0}{|x^A_{12}|^2}\Bigg) \right. \nn
&\;\; +∫dτ_3∫d^3x_0\frac{\dot{x}_{A,1}^{αβ}\dot{x}_{A,2}^{γδ}\dot{x}_{A,3}^{κλ}(x_{01})_{βγ}(x_{02})_{δκ}(x_{03})_{λα} -\O({θ_i}^2)}{4·16π^3|x_{01}|^3|x_{02}|^3|x_{03}|^3}\Bigg\}.
\label{finres}}

\section{Comments}\label{Sc:com}
We have constructed a $\frac{1}{3}$-BPS supersymmetrized Wilson loop operator in $d=3$, $\N=3$ harmonic superspace for CS theories. This operator readily generalizes the $\frac{1}{6}$-BPS operator for ABJM theories. We were also able to use the power of harmonic superspace to compute the `one-loop' perturbative corrections directly in superspace.

Using the component expansion of $\N=3$ connections and field strengths, and just focussing on the localization locus discussed in Section \ref{Sc:Wu} ($\s≡φ^{0}$ and $D≡X^{0}=-\frac{\s}{r}$), we can see that the $\W_{\sfrac{1}{3}}$ given in \eqref{WL13s} reduces to \eqref{WLN2} once we identify $θ,\bar{θ}$ with $θ^{++},θ^{--}$. Then one can expect that
\equ{\braket{\W_{\sfrac{1}{3}}}=1 +\frac{1}{2}\left[ϑ+\frac{ϑ^2}{4}\right]α -\left[\frac{1}{24}\(5+\frac{1}{N^2}\) +\frac{1}{4}ϑ -\frac{1}{6}\(\frac{1}{2}+\frac{1}{N^2}\)\frac{ϑ^2}{4}\right]α²+\O(α^3).
\label{locres}}
At order $α^2$, we can directly compare the `bosonic' factor $-\frac{5α^2}{24}≡\frac{5π^2N^2}{6k^2}$ above to the corresponding perturbative expression in \eqref{finres}. They exactly match once we perform the integrals in the latter case for a circular WL, i.e., $x^μ=\(0,\sin(τ),\cos(τ)\)$ as one might expect\footnote{We refer the readers to \cite{Rey:2008bh} for evaluation of the relevant integrals.}.

Formally, both \eqref{finres} and \eqref{locres} have nonvanishing `fermionic' contributions at $\O(α)$ and $\O(α^2)$. However, without knowing the explicit profile functions of $θ(τ)$ and $u(τ)$ we cannot proceed further. However, we can identify a contribution at $\O(α^2)$ that does not receive any $\O(\frac{1}{N})$ corrections!\footnote{We do not have $\O(\frac{1}{N})$ terms at $\O(α)$ either, but that could still be treated as a phase factor. The striking feature of the term at $\O(α^2)$ is that it remains unchanged even after the removal of the $\O(α)$ phase:
\[\braket{\W_{\sfrac{1}{3}}}=1 -\left[\frac{1}{24}\(5+\frac{1}{N^2}\) +\frac{1}{4}ϑ +\frac{1}{6}\(\frac{5}{2}-\frac{1}{N^2}\)\frac{ϑ^2}{4}\right]α²+\O(α^3).\]
} These are the $\O(θ\bar{θ})$ terms in the $ϑ$ piece of \eqref{locres} and comparing with \eqref{finres}, we can give an explicit expression for these terms:
\eqst{\left.ϑ\right|_{\O(θ\bar{θ})}=\frac{2\i}{π^2}∫dτ_1dτ_2\frac{\big(\dot{x}_1·\dot{x}_2 -|\dot{x}_1||\dot{x}_2|\big)x_{12}^{αβ}\big(θ_{1,α}\bar{θ}_{1,β} -θ_{2,α}\bar{θ}_{2,β}\big)}{|x_{12}|^4} \\
-\frac{\i}{16π^3}∫dτ_{1,2,3}\,\dot{x}_1^{αβ}\dot{x}_2^{γδ}\dot{x}_3^{κλ}∫d^3x_0\frac{(x_{01})_{βγ}(x_{02})_{δκ}θ_{3,λ}\bar{θ}_{3,α} +\text{2 more terms}}{|x_{01}|^3|x_{02}|^3|x_{03}|^3}\,·
\label{conject}}
The fermionic pieces from the term proportional to $\big(\dot{x}_{A,1}·\dot{x}_{A,2} -|\dot{x}_{A,1}||\dot{x}_{A,2}|\big)$ do not contribute above because the combination $θ\bar{θ}$ is independent of $τ$ as discussed in Section \ref{Sc:Wu}. Such fermionic contributions to the Wilson loop operators do not seem to have been considered in $d=3$ but similar terms have appeared in the $d=4$, $\N=4$ SYM literature, specifically in the study of supersymmetrized Maldacena-Wilson loops\cite{Muller:2013rta,Beisert:2015jxa}. Thus, a careful study of the $τ$-dependence of the $θ$ and $u$ coordinates that is consistent with the `bosonic' circular WL is required to understand how the general perturbative result \eqref{finres} reduces to the simpler localization result \eqref{locres} at various $θ$ orders.\footnote{This is the case when the conjectured equality \eqref{conject} would hold and we assume that the consistency would require $θ^0(τ)$ to vanish.} We leave this exercise for future work.

One can also ask whether the construction of a $\half$-BPS WL with `supermatrix' structure\cite{Ouyang:2015bmy} is feasible in harmonic superspace. Our preliminary analysis suggests that the $U(1)$ charge structure of the supersymmetry parameters $(ε^{±±},ε^0)$ and the matter superfield $q^+$ is an obstruction for constructing a straightforward generalization. As mentioned in the Introduction, a motivation to study such supersymmetrized WL operators is to probe Wilson loops/Scattering amplitudes duality in ABJM theory. The expectation is that polygonal WL operators with certain bifundamental vertex insertions would be dual to the ABJM scattering amplitudes. The matter superfield $(q^{+a})^B_A$ in bifundamental representation provides a natural candidate for such insertions. However, this leads to some superficial divergences that need to be tamed. Progress on these aspects will be reported elsewhere.

\section*{\centering Acknowledgements}
DJ thanks Kazuo Hosomichi for insightful discussions spanning both conceptual and technical details found here. DJ also thanks Warren Siegel for helpful discussions regarding Wilson loops and $♫/\check{Π}$ superspaces. YtH is supported in part by MoST Grant No. 106-2628-M-002-012-MY3.

\references{
\bibitem{Drummond:2007cf}
J.~M.~Drummond, J.~Henn, G.~P.~Korchemsky and E.~Sokatchev,
``On Planar Gluon Amplitudes/Wilson Loops Duality'',
\npb{795}{2008}{52}{10.1016/j.nuclphysb.2007.11.007} [\arXivid{0709.2368}{hep-th}].

\bibitem{Drummond:2007bm}
J.~M.~Drummond, J.~Henn, G.~P.~Korchemsky and E.~Sokatchev,
``The Hexagon Wilson Loop and the BDS Ansatz for the Six-Gluon Amplitude'',
\plb{662}{2008}{456}{10.1016/j.physletb.2008.03.032} [\arXivid{0712.4138}{hep-th}].

\bibitem{Bern:2008ap}
Z.~Bern, L.~J.~Dixon, D.~A.~Kosower, R.~Roiban, M.~Spradlin, C.~Vergu and A.~Volovich,
``The Two-Loop Six-Gluon MHV Amplitude in Maximally Supersymmetric Yang-Mills Theory'',
\prd{78}{2008}{045007}{10.1103/PhysRevD.78.045007} [\arXivid{0803.1465}{hep-th}].

\bibitem{Drummond:2008aq}
J.~M.~Drummond, J.~Henn, G.~P.~Korchemsky and E.~Sokatchev,
``Hexagon Wilson Loop = Six-Gluon MHV Amplitude'',
\npb{815}{2009}{142}{10.1016/j.nuclphysb.2009.02.015} [\arXivid{0803.1466}{hep-th}].

\bibitem{Drummond:2008vq}
J.~M.~Drummond, J.~Henn, G.~P.~Korchemsky and E.~Sokatchev,
``Dual Superconformal Symmetry of Scattering Amplitudes in $\N=4$ Super-Yang-Mills Theory'',
\npb{828}{2010}{317}{10.1016/j.nuclphysb.2009.11.022} [\arXivid{0807.1095}{hep-th}].

\bibitem{Berkovits:2008ic}
N.~Berkovits and J.~Maldacena,
``Fermionic T-Duality, Dual Superconformal Symmetry, and the Amplitude/Wilson Loop Connection'',
\jhep{0809}{2008}{062}{10.1007/JHEP09(2008)062} [\arXivid{0807.3196}{hep-th}].

\bibitem{Mason:2010yk}
L.~J.~Mason and D.~Skinner,
``The Complete Planar S-matrix of $\N=4$ SYM as a Wilson Loop in Twistor Space'',
\jhep{1012}{2010}{018}{10.1007/JHEP12(2010)018} [\arXivid{1009.2225}{hep-th}].

\bibitem{CaronHuot:2010ek}
S.~Caron-Huot,
``Notes on the Scattering Amplitude/Wilson Loop Duality'',
\jhep{1107}{2011}{058}{10.1007/JHEP07(2011)058} [\arXivid{1010.1167}{hep-th}].

\bibitem{Aharony:2008ug}
O.~Aharony, O.~Bergman, D.~L.~Jafferis and J.~Maldacena,
``$\N=6$ Superconformal Chern-Simons-Matter Theories, M2-branes and Their Gravity Duals'',
\jhep{0810}{2008}{091}{10.1007/JHEP10(2008)091} [\arXivid{0806.1218}{hep-th}].

\bibitem{Chen:2011vv}
W.~M.~Chen and Y.~t.~Huang,
``Dualities for Loop Amplitudes of $\N=6$ Chern-Simons Matter Theory'',
\jhep{1111}{2011}{057}{10.1007/JHEP11(2011)057} [\arXivid{1107.2710}{hep-th}].

\bibitem{Bianchi:2011dg}
M.~S.~Bianchi, M.~Leoni, A.~Mauri, S.~Penati and A.~Santambrogio,
``Scattering Amplitudes/Wilson Loop Duality in ABJM Theory'',
\jhep{1201}{2012}{056}{10.1007/JHEP01(2012)056} [\arXivid{1107.3139}{hep-th}].

\bibitem{Gaiotto:2007qi}
D.~Gaiotto and X.~Yin,
``Notes on Superconformal Chern-Simons-Matter Theories'',
\jhep{0708}{2007}{056}{10.1007/JHEP08(2007)056} [\arXivid{0704.3740}{hep-th}].

\bibitem{Drukker:2008zx}
N.~Drukker, J.~Plefka and D.~Young,
``Wilson Loops in 3-dimensional $\N=6$ Supersymmetric Chern-Simons Theory and Their String Theory Duals'',
\jhep{0811}{2008}{019}{10.1007/JHEP11(2008)019} [\arXivid{0809.2787}{hep-th}].

\bibitem{Chen:2008bp}
B.~Chen and J.~B.~Wu,
``Supersymmetric Wilson Loops in $\N=6$ Super Chern-Simons-Matter Theory'',
\npb{825}{2010}{38}{10.1016/j.nuclphysb.2009.09.015} [\arXivid{0809.2863}{hep-th}].

\bibitem{Rey:2008bh}
S.~J.~Rey, T.~Suyama and S.~Yamaguchi,
``Wilson Loops in Superconformal Chern-Simons Theory and Fundamental Strings in Anti-de Sitter Supergravity Dual'',
\jhep{0903}{2009}{127}{10.1007/JHEP03(2009)127} [\arXivid{0809.3786}{hep-th}].

\bibitem{Kapustin:2009kz}
A.~Kapustin, B.~Willett and I.~Yaakov,
``Exact Results for Wilson Loops in Superconformal Chern-Simons Theories with Matter'',
\jhep{1003}{2010}{089}{10.1007/JHEP03(2009)127} [\arXivid{0909.4559}{hep-th}].

\bibitem{Drukker:2009hy}
N.~Drukker and D.~Trancanelli,
``A Supermatrix Model for $\N=6$ Super Chern-Simons-Matter Theory'',
\jhep{1002}{2010}{058}{10.1007/JHEP02(1010)058} [\arXivid{0912.3006}{hep-th}].

\bibitem{Bianchi:2013zda}
M.~S.~Bianchi, G.~Giribet, M.~Leoni and S.~Penati,
``1/2 BPS Wilson Loop in $\N=6$ Superconformal Chern-Simons Theory at Two Loops'',
\prd{88}{2013}{026009}{10.1103/PhysRevD.88.026009} [\arXivid{1303.6939}{hep-th}].

\bibitem{Bianchi:2013iha}
M.~S.~Bianchi, M.~Leoni, M.~Leoni, A.~Mauri, S.~Penati and A.~Santambrogio,
``ABJM Amplitudes and WL at Finite $N$'',
\jhep{1309}{2013}{114}{10.1007/JHEP09(2013)114} [\arXivid{1306.3243}{hep-th}].

\bibitem{Ouyang:2015bmy}
H.~Ouyang, J.~B.~Wu and J.~j.~Zhang,
``Construction and Classification of Novel BPS Wilson Loops in Quiver Chern-Simons-Matter Theories'',
\npb{910}{2016}{496}{10.1016/j.nuclphysb.2016.07.018} [\arXivid{1511.02967}{hep-th}].

\bibitem{Mauri:2017whf}
A.~Mauri, S.~Penati and J.~j.~Zhang,
``New BPS Wilson Loops in $\N=4$ Circular Quiver Chern-Simons-Matter Theories'',
\jhep{1711}{2017}{174}{10.1007/JHEP11(2017)174} [\arXivid{1709.03972}{hep-th}].

\bibitem{Rosso:2014oha}
M.~Rosso and C.~Vergu,
``Wilson Loops in $\N=6$ Superspace for ABJM Theory'',
\jhep{1406}{2014}{176}{10.1007/JHEP06(2014)176} [\arXivid{1403.2336}{hep-th}].

\bibitem{Beisert:2012gb}
N.~Beisert and C.~Vergu,
``On the Geometry of Null Polygons in Full $\N=4$ Superspace'',
\prd{86}{2012}{026006}{10.1103/PhysRevD.86.026006} [\arXivid{1203.0525}{hep-th}].

\bibitem{Buchbinder:2008vi}
I.~L.~Buchbinder, E.~A.~Ivanov, O.~Lechtenfeld, N.~G.~Pletnev, I.~B.~Samsonov and B.~M.~Zupnik,
``ABJM Models in $\N=3$ Harmonic Superspace'',
\jhep{0903}{2009}{096}{10.1007/JHEP03(2009)096} [\arXivid{0811.4774}{hep-th}].

\bibitem{Buchbinder:2009dc}
I.~L.~Buchbinder, E.~A.~Ivanov, O.~Lechtenfeld, N.~G.~Pletnev, I.~B.~Samsonov and B.~M.~Zupnik,
``Quantum $\N=3$, $d=3$ Chern-Simons Matter Theories in Harmonic Superspace'',
\jhep{0910}{2009}{075}{10.1007/JHEP10(2009)075} [\arXivid{0909.2970}{hep-th}].

\bibitem{Galperin:1984av}
A.~Galperin, E.~Ivanov, S.~Kalitsyn, V.~Ogievetsky and E.~Sokatchev,
``Unconstrained $\N=2$ Matter, Yang-Mills and Supergravity Theories in Harmonic Superspace'',
\cqg{1}{1984}{469}{10.1088/0264-9381/1/5/004}; Erratum: \cqg{2}{1985}{127}{10.1088/0264-9381/2/1/512}.

\bibitem{Nishino:1991sr}
H.~Nishino and S.~J.~Gates, Jr.,
``Chern-Simons Theories with Supersymmetries in Three-dimensions'',
\href{https://doi.org/10.1142/S0217751X93001363}{{\it Int. J. Mod. Phys. A} {\bf 08} (1993) 3371}.

\bibitem{Gates:1991qn}
S.~J.~Gates, Jr. and H.~Nishino,
``Remarks on the $\N=2$ Supersymmetric Chern-Simons Theories'',
\plb{281}{1992}{72}{10.1016/0370-2693(92)90277-B}.

\bibitem{Beisert:2015jxa}
N.~Beisert, D.~Müller, J.~Plefka and C.~Vergu,
``Smooth Wilson Loops in $\N=4 $ Non-chiral Superspace'',
\jhep{1512}{2015}{140}{10.1007/JHEP12(2015)140} [\arXivid{1506.07047}{hep-th}].

\bibitem{Galperin:2001uw}
A.~S.~Galperin, E.~A.~Ivanov, V.~I.~Ogievetsky and E.~S.~Sokatchev,
{\it Harmonic Superspace}, \href{https://doi.org/10.1017/CBO9780511535109}{Cambridge University Press, Cambridge, UK (2001) 306p}.

\bibitem{Muller:2013rta}
D.~Müller, H.~Münkler, J.~Plefka, J.~Pollok and K.~Zarembo,
``Yangian Symmetry of Smooth Wilson Loops in $\N=4$ Super Yang-Mills Theory'',
\jhep{1311}{2013}{081}{10.1007/JHEP11(2013)081} [\arXivid{1309.1676}{hep-th}].
}

\end{document}